\documentclass[twocolumn,preprintnumbers,amsmath,amssymb,floatfix,prl]{revtex4}
\usepackage{graphicx}
\usepackage{dcolumn}
\usepackage{bm}
\usepackage[usenames,dvipsnames]{color}% use as \textcolor{red}{******} 

\begin{document}

\title{Strong Coupling Nature of the Excitonic Insulator State in Ta$_2$NiSe$_5$}

\author{Koudai Sugimoto$^1$}
\author{Satoshi Nishimoto$^{2,3}$}
\author{Tatsuya Kaneko$^4$}
\author{Yukinori Ohta$^5$}
\affiliation{$^1$Center for Frontier Science, Chiba University, Chiba 263-8522, Japan}
\affiliation{$^2$Department of Physics, Technical University Dresden, 01069 Dresden, Germany}
\affiliation{$^3$Institute for Theoretical Solid State Physics, IFW Dresden, 01171 Dresden, Germany}
\affiliation{$^4$Computational Condensed Matter Physics Laboratory, RIKEN, Wako, Saitama 351-0198, Japan}
\affiliation{$^5$Department of Physics, Chiba University, Chiba 263-8522, Japan}

\date{\today}

\begin{abstract}
We analyze the measured optical conductivity spectra using the density-functional-theory-based electronic 
structure calculation and density-matrix renormalization group calculation of an effective model.  We show 
that, in contrast to a conventional description, the Bose-Einstein condensation of preformed excitons occurs 
in Ta$_2$NiSe$_5$, despite the fact that a noninteracting band structure is a band-overlap semimetal rather 
than a small band-gap semiconductor.  The system above the transition temperature is therefore not 
a semimetal but rather a state of preformed excitons with a finite band gap.  A novel insulator state 
caused by the strong electron-hole attraction is thus established in a real material.  
\end{abstract}

%\pacs{}

\maketitle

%%%%%%%%%%%%%%%
%\section{Introduction}
%%%%%%%%%%%%%%%

\begin{figure}[tb]
\begin{center}
\includegraphics[width=0.7\columnwidth]{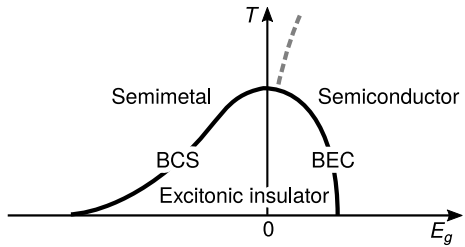}
\end{center}
\caption{Schematic representation of the conventional phase diagram of an excitonic insulator 
\cite{Kozlov1965JETP}.  The band gap $E_g$ in the noninteracting band structure is either 
positive (semiconducting) or negative (semimetallic).  The crossover between the semimetallic and 
semiconducting states at finite temperatures ($T$) is indicated by the dashed line.  
}\label{fig1}
\end{figure}

It has been commonly understood \cite{Mott1961PM, Jerome1967PR, Halperin1968RMP, Kunes2015JPCM} 
that excitonic condensation occurs when electrons and holes in a small band-overlap semimetal 
or a small band-gap semiconductor form pairs (or excitons) owing to the attractive Coulomb interaction 
and condense into a quantum state with macroscopic phase coherence. The excitonic condensation 
in semimetallic systems can be described in analogy with the BCS theory of superconductors,
and that in semiconducting systems can be discussed in terms of the Bose-Einstein condensation 
(BEC) of preformed excitons \cite{Bronold2006PRB,Seki2011PRB,Zenker2012PRB}.  
A well-known phase diagram \cite{Kozlov1965JETP} is depicted in Fig.~\ref{fig1}.  

In this Letter, we will show that this conventional picture established half a century ago \cite{Kozlov1965JETP} 
is, in fact, seriously violated in a candidate material Ta$_2$NiSe$_5$ recently found 
\cite{Wakisaka2009PRL,Kaneko2013PRB,Seki2014PRB,Kim2016ACSNANO,Lu2017NC}.  Namely, based on the 
analyses of the measured optical conductivity spectra \cite{Larkin2016Thesis, Lu2017NC, Larkin2017PRB}, 
we will show that the excitonic BEC occurs in Ta$_2$NiSe$_5$, despite the fact that the 
noninteracting band structure is a band-overlap semimetal rather than a small band-gap 
semiconductor.  In other words, even though the noninteracting band structure is semimetallic, 
the system above the transition temperature ($T_c$) is not a semimetal but rather a state of 
preformed excitons with a finite band gap, exhibiting a variety of intriguing physical properties.  
A novel insulator state caused by the strong electron-hole attraction is thus established, 
which may be contrasted with a Mott insulator state caused by the strong electron-electron repulsion.

%%%%%%%%%%%%%%%
%\section{Brief history}
%%%%%%%%%%%%%%%

Let us first review essential physical properties of Ta$_2$NiSe$_5$ briefly. 
This material has a layered structure stacked loosely by a weak van der Waals interaction, and 
in each layer, Ni single chains and Ta double chains run along the $a$ axis of the lattice 
to form a quasi-one-dimensional (1D) chain structure \cite{Sunshine1985IC}.  
The observed resistivity shows a semiconducting behavior over a wide temperature range up to 
$\sim 500$~K \cite{DiSalvo1986JLCM}, with a quasi-1D anisotropic electron conduction \cite{Lu2017NC}.  
An anomaly in the resistivity appears at $T_c \simeq 328$~K, which is associated with a second-order 
structural phase transition from the orthorhombic to the monoclinic phase \cite{DiSalvo1986JLCM}.  
The magnetic susceptibility exhibits diamagnetism in a wide temperature range and shows a 
sudden drop (being more negative) below $T_c$ \cite{DiSalvo1986JLCM}.  
The angle-resolved photoemission spectroscopy (ARPES) experiment \cite{Wakisaka2009PRL,Seki2014PRB} 
showed that, by lowering the temperature, the flatness of the top of the valence band is enhanced and the size 
of the band gap becomes wider, thereby suggesting that the excitonic insulator state is realized as 
the ground state of this material \cite{Wakisaka2009PRL}.  

Kaneko \textit{et al.} \cite{Kaneko2013PRB} then made the density-functional-theory (DFT) -based 
electronic structure calculations for the orthorhombic phase of Ta$_2$NiSe$_5$ and found that the system 
is a direct-gap semiconductor with the gap minimum at the $\Gamma$ point of the Brillouin zone.  
No hybridization occurs between the top of the valence band and the bottom of the conduction band, 
which belong to different irreducible representations at the $\Gamma$ point.  
The effective three-chain Hubbard model was then constructed to reproduce three bands near 
the Fermi level.  This model, together with the phonon degrees of freedom \cite{Kaneko2015PRB}, was 
analyzed by the mean-field approximation, and the ground-state and finite-temperature phase diagrams 
were obtained to show that the BEC of excitons cooperatively induces the structural phase transition \cite{Kaneko2013PRB}. 
The spontaneous hybridization between the conduction and valence bands well reproduces the 
flattening of the top of the valence band observed in the ARPES experiment \cite{Kaneko2013PRB}.  
A number of physical quantities have also been discussed along this line \cite{Sugimoto2016PRB, Yamada2016JPSJ, Sugimoto2016PRB2, Matsuura2016JPSJ}.  

%%%%%%%%%%%%%%%
%\section{Optical conductivity leading to reconsideration of the electronic state of Ta2NiSe5}
%%%%%%%%%%%%%%%

Recently, a comparative experimental study of the optical conductivity spectra of Ta$_2$NiSe$_5$ 
and Ta$_2$NiS$_5$ was made \cite{Larkin2017PRB,Larkin2016Thesis}, which provided us with an 
unexpected opportunity to reconsider the validity of this conventional description.  
The latter material Ta$_2$NiS$_5$ is isostructural to the high-temperature phase of Ta$_2$NiSe$_5$ 
but shows no structural phase transition, which may therefore be regarded as a simple semiconductor 
with a much larger band gap ($\sim 0.25$ eV at 150 K) than that of Ta$_2$NiSe$_5$ ($\sim 0.16$ eV) 
\cite{Larkin2017PRB}.  
Most notably, a huge peak appears at $\sim0.4$ eV in the optical conductivity spectrum of 
Ta$_2$NiSe$_5$ in a wide temperature range both below and above $T_c$ \cite{Lu2017NC}, which 
is absent in the spectrum of Ta$_2$NiS$_5$.  We consider the origin of this peak using two approaches:
a DFT-based electronic structure calculation and a density-matrix renormalization group (DMRG) 
calculation~\cite{DMRG} of an effective model.  As we will discuss below, these calculations lead 
us to the remarkable conclusion given briefly above.  

%%%%%%%%%%%%%%%
%\section{Method of DFT calculations}
%%%%%%%%%%%%%%%

We carry out the DFT-based electronic structure calculations using the WIEN2k code \cite{WIEN2k}, 
where we use the generalized gradient approximation for electron correlations with the exchange 
correlation potential given in Ref.~[\onlinecite{Perdew1996PRL}].  
The crystal structures of the high-temperature orthorhombic phase (space group $Cmcm$) are taken 
from Ref.~[\onlinecite{JainAPLM2013}].
Because the DFT-based band calculations usually underestimate the band gap, we introduce 
a modified Becke-Johnson (MBJ) exchange potential \cite{Tran2009PRL, Koller2011PRB} 
with a parameter $c$ defined in Eq.~(3) of Ref.~[\onlinecite{Tran2009PRL}] and improve this 
underestimation.  We thereby find that the band gap opens at $c > 1.2$ for Ta$_2$NiS$_5$ but an 
unusually large value of $c > 1.63$ is required to open the band gap for Ta$_2$NiSe$_5$.  
We choose a value $c = 1.5$ in the following calculations \cite{Note1}.  
Details of our band calculations are given in Supplemental Material \cite{SM}.  

%%%%%%%%%%%%%%%
%\section{Results of DFT calculations and comparison with experiment}
%%%%%%%%%%%%%%%

\begin{figure}[tb]
\begin{center}
\includegraphics[width=0.95\columnwidth]{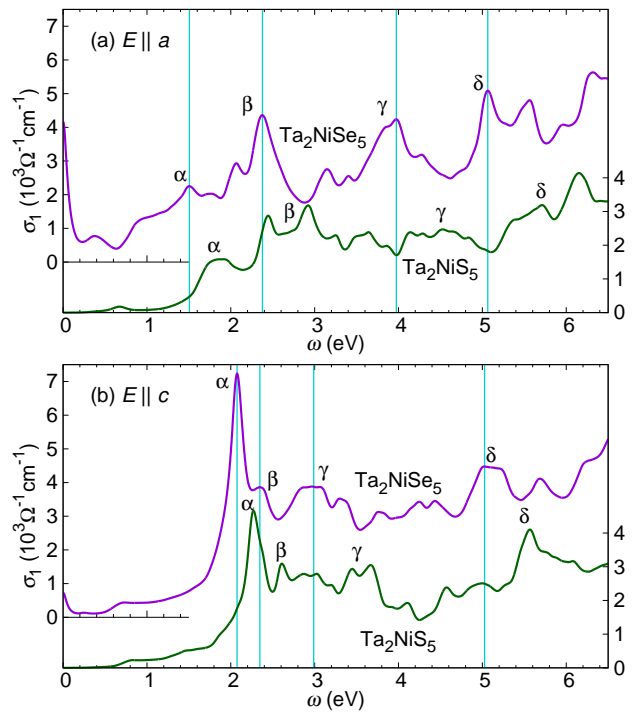}
\end{center}
\caption{Calculated optical conductivity spectra of Ta$_2$NiSe$_5$ (purple line) and Ta$_2$NiS$_5$ 
(green line) for the electric field $E$ parallel to (a) the crystallographic $a$ axis and (b) the $c$ axis.  
Vertical lines indicate the positions of the major peaks for Ta$_2$NiSe$_5$, which are labeled 
$\alpha$, $\beta$, $\gamma$, and $\delta$ following Ref.~[\onlinecite{Larkin2017PRB}] in (a).  
The same labels are put in (b) but have no relation to the labels in (a).  
}\label{fig2}
\end{figure}

In the calculation of the optical conductivity spectra \cite{Draxl2006CPC}, we use 5046 
$k$ points in the irreducible part of the Brillouin zone, and a broadening parameter of the spectra 
is set to be 0.05 eV for both interband and intraband (Drude) contributions.  
The real part of the optical conductivity spectra $\sigma_1(\omega)$ for the electric field $E$ 
parallel to the $a$ and $c$ axes is shown in Figs.~\ref{fig2}(a) and \ref{fig2}(b), respectively, for both 
Ta$_2$NiSe$_5$ and Ta$_2$NiS$_5$.  
The results for $E\parallel a$ should be compared with experimental data given in Fig.~1 of 
Ref.~[\onlinecite{Larkin2017PRB}], where the same labeling of the peaks $\alpha$, $\beta$, $\gamma$, 
and $\delta$ are used, and the results for $E\parallel c$ should be compared with experimental 
data given in Figs.~4.9 and 4.10 of Ref.~[\onlinecite{Larkin2016Thesis}].  
We first find in Fig.~\ref{fig2}(b) that the peak at $\omega \simeq 2$ eV in the spectra of 
$E \parallel c$ is well reproduced by our calculations for both Ta$_2$NiSe$_5$ and Ta$_2$NiS$_5$ 
if we assume $c = 1.5$.  The higher-energy spectral features are also well reproduced.  
We moreover find in Fig.~\ref{fig2}(a) that the series of peaks at $\omega\agt 1.2$ eV in the spectra 
of $E\parallel a$ labeled $\alpha$, $\beta$, $\gamma$, and $\delta$ are all well reproduced by 
our calculations with the same value $c = 1.5$.  
These agreements with experiment are remarkable in that the DFT-based calculations using 
the MBJ potential with an appropriate choice of the $c$ value can reproduce all the high-energy 
spectral features very well in both materials \cite{Note2}.

However, we notice in Fig.~\ref{fig2} that this approach is quite poor in the description of the 
low-energy region $\omega\alt 1.1$~eV, in particular for Ta$_2$NiSe$_5$; i.e., the calculation 
predicts a metallic state although insulating in the experiment, and a huge peak observed at 
$\omega \simeq 0.4$~eV \cite{Lu2017NC} is absent in the calculated spectrum of $E \parallel a$.  
If Ta$_2$NiSe$_5$ above $T_c$ is a simple semiconductor, just as Ta$_2$NiS$_5$, the low-energy 
peak should not appear because the dipole transition at the lowest-energy region is prohibited \cite{Note3}.  
On the other hand, if Ta$_2$NiSe$_5$ is an excitonic insulator of the BCS type, it should be 
metallic above $T_c$, which is not consistent with the experiment either.  The origin of this peak 
is thus raised as an important issue, for which we need to invent the key to a solution.  
Larkin \textit{et al.}~\cite{Larkin2017PRB} attributed the origin of this peak to the giant 
oscillator strength of spatially extended exciton-phonon bound states.  
Below, we will present a different solution.  

%%%%%%%%%%%%%%%
%\subsection{Low-energy spectral features and DMRG calculation of the model}
%%%%%%%%%%%%%%%

In order to consider this issue further, we adopt the 1D extended Falicov-Kimball model (a spinless 
two-band model), a minimum lattice model to discuss the spin-singlet excitonic condensation 
\cite{Seki2014PRB,Ejima2014PRL}.  
We thereby calculate the optical conductivity spectrum, as well as the single-particle spectrum, to 
consider the low-energy excitations of the model and origin of the peak in the optical conductivity 
spectrum.  The Hamiltonian reads 
\begin{align}
\cal{H} =&
	-t_c \sum_{\langle i, j \rangle} \left( c^{\dagger}_{i} c_j + \mathrm{H.c.} \right)
	-t_f \sum_{\langle i, j \rangle} \left( f^{\dagger}_{i} f_j + \mathrm{H.c.} \right) 
\notag \\
	&
	+ \frac{D}{2} \sum_{i} \left( c^{\dagger}_{i} c_i - f^{\dagger}_{i} f_i \right)
        + V \sum_{i} c^{\dagger}_{i} c_i f^{\dagger}_{i} f_i,
\end{align}
where $c^{\dagger}_{i}$ ($f^{\dagger}_{i}$) creates an electron in the conduction (valence) band 
orbital at site $i$, $\langle i, j \rangle$ indicates the nearest-neighbor pair of sites,  
$t_c$ and $t_f$ are the hopping integrals of $c$ and $f$ orbitals, respectively, assuming a direct 
band gap ($t_ct_f<0$), $D$ is the level separation between $c$ and $f$ orbitals, and $V$ 
is the interorbital Coulombic repulsion.  We consider the case at half filling.  
We apply the dynamical DMRG method \cite{DDMRG} to calculate the spectra, which are given by 
the imaginary part of the dynamical correlation functions 
\begin{equation}
I(\omega)=\frac{1}{\pi} \mathrm{Im} \langle \psi_0|A^\dagger\frac{1}{{\cal H}+\omega-E_0-i \eta}A|\psi_0\rangle,
\end{equation}
where $E_0$ and $|\psi_0\rangle$ are the ground-state energy and wave function, respectively, 
and $A$ is the quantum operator corresponding to the physical quantity analyzed.  
A small real number $\eta$ $(>0)$ is introduced in the calculation to shift the poles of the correlation 
function into the complex plane.  
For the optical conductivity spectrum, we employ the dipole operator 
$d=-e\sum_{l=1}^L l(c^\dagger_l c_l + f^\dagger_l f_l - 1)$ for $A$. Then, the real part of the optical 
conductivity $\sigma_1$ is obtained as $\sigma(\omega)=\omega I(\omega)$. The deconvolution 
technique \cite{deconv} is used to obtain the spectrum in the thermodynamic limit $\eta\to0$.  
For the single-particle spectrum, we simply take either $A=c_k$ $(f_k)$ or 
$A =c^\dagger_k$ $(f^\dagger_k)$ to calculate the photoemission or inverse photoemission spectra 
for the conduction (valence) band orbital.  Since open boundary conditions are applied here, 
the momentum-dependent operators are defined as $c_k=\sqrt{2/(L+1)}\sum_l\sin(kl)c_l$ with 
quasimomentum $k=\pi z/(L+1)$ for integers $1 \le z \le L$.

\begin{figure}[tb]
\begin{center}
\includegraphics[width=0.95\columnwidth]{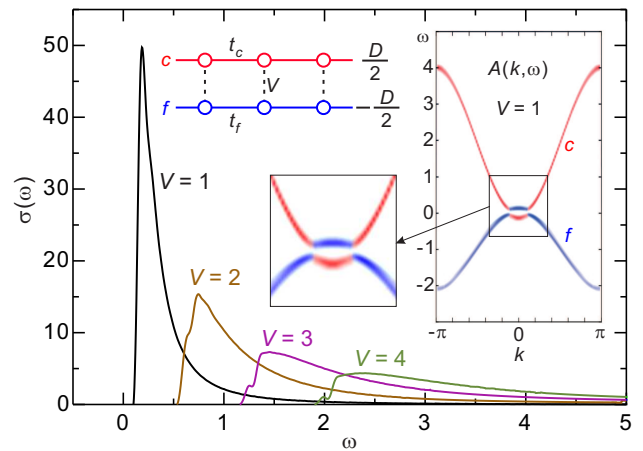}
\end{center}
\caption{Calculated optical conductivity spectra $\sigma(\omega)$ of the extended Falicov-Kimball 
model using an $L\times2=100\times2$ cluster.  We assume $t_c=1$ and $t_f=-0.5$, and $D$ is 
chosen to keep $\langle c^\dagger_i c_i \rangle=0.1$ for each $V$.  The inset shows the calculated 
single-particle spectrum $A(k,\omega)$ of the same model with $L\times2=60\times2$ and its 
enlargement near the Fermi level, where the broadening is $\eta=0.05$.  
}\label{fig3}
\end{figure}

The calculated results are shown in Fig.~\ref{fig3}, where we find that a large asymmetric peak 
appears in the optical conductivity spectrum $\sigma(\omega)$.  The peak position shifts upward 
and the peak width broadens with increasing $V$.  
Intuitively, we can understand the presence of the peak as follows:  Assume a semimetallic case 
in the strong-coupling limit; then the electron in the conduction band, when propagates, necessarily 
interacts with the electron in the valence band, and the energy of the conduction-band electron 
increases by $V$, giving rise to a peak at the energy $\omega\simeq V$.  The electron-electron 
repulsive interaction $V$ acts as an electron-hole attractive interaction, so that $V$ gives the 
exciton binding energy.  The band gap of the size $\sim V$ also opens in the single-particle 
spectrum.  Thus, the presence of the interorbital interaction is enough to yield the peak in the 
optical conductivity spectrum, as well as the band gap in the single-particle spectrum, 
irrespective of whether the long-range order of the BEC occurs or not.  
The energy of the peak and the size of the band-gap scale with $V$ in the strong-coupling region, 
although they are much reduced in the intermediate- to weak-coupling region as shown in 
Fig.~\ref{fig3}.  We should stress that this situation occurs only when the noninteracting band structure 
is semimetallic; when it is semiconducting, the peak vanishes, irrespective of $V$, unless spontaneous
$c$-$f$ hybridization occurs.  We then immediately notice that the semimetallic 
situation in our model corresponds to Ta$_2$NiSe$_5$ and the semiconducting situation in our 
model corresponds to Ta$_2$NiS$_5$.

This explanation of the presence of the peak in the optical conductivity spectrum, together with 
the opening of the band gap in the semimetallic noninteracting band structure, is very simple and 
straightforward, which we believe should be applicable to the low-energy physics of Ta$_2$NiSe$_5$, 
although more quantitative calculations assuming, e.g., a three-chain model with electron-phonon 
coupling \cite{Kaneko2015PRB}, would give an improved description.  

%%%%%%%%%%%%%%%
%\subsection{Mean-field phase diagrams and its implication: a novel insulator!! }
%%%%%%%%%%%%%%%

\begin{figure}[tb]
\begin{center}
\includegraphics[width=0.95\columnwidth]{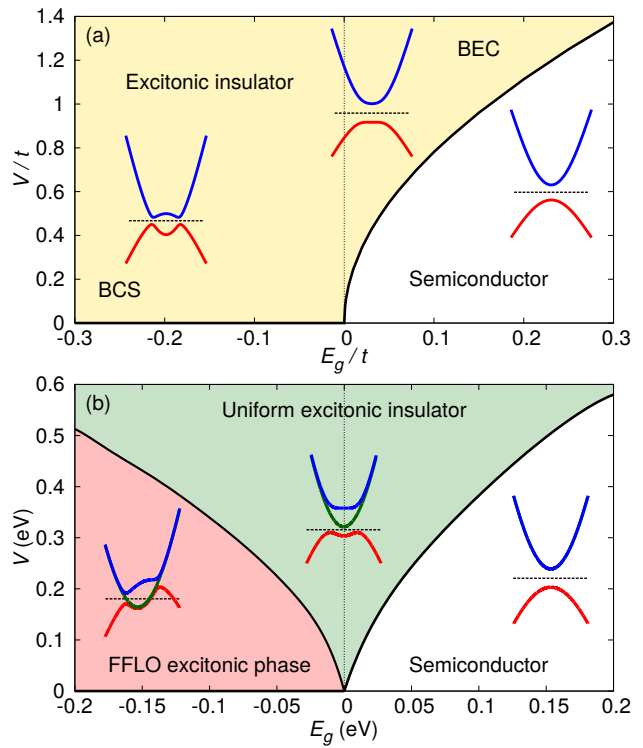}
\end{center}
\caption{Ground-state phase diagrams of 
(a) the two-chain Hubbard model with $t_c=t$, $t_f=-0.5t$, and $U=2V$ and 
(b) the three-chain Hubbard model with $t_c = 0.8$~eV, $t_f = -0.4$~eV, and $U=4V$ 
calculated by the mean-field approximation, neglecting the Hartree shift \cite{SM}.  
Corresponding quasiparticle band dispersions are also illustrated schematically.  
}\label{fig4}
\end{figure}

To envisage the excitonic insulator states in the entire parameter space, we 
present the ground-state phase diagrams of the two-chain and three-chain Hubbard 
models calculated in the mean-field approximation 
\cite{Kaneko2013PRB,Sugimoto2016PRB,Yamada2016JPSJ,SM}.
The results in the spin-singlet channel are shown in Fig.~\ref{fig4} in the parameter space 
of the level separation [or noninteracting band gap $E_g=D-2(|t_c|+|t_f|)$] and the interorbital 
repulsive interaction $V$.  The corresponding quasiparticle band dispersions are also illustrated.  
For the two-chain Hubbard model [see Fig.~\ref{fig4}(a)], we find that the ground state in 
the weak-coupling regime is either a band insulator when $E_g>0$ or an excitonic insulator 
when $E_g<0$, just as the conventional phase diagram indicates (see Fig.~\ref{fig1}).    
In the strong-coupling regime, on the other hand, the excitonic insulator state of the 
BEC type comes out over a wide region around $E_g \simeq 0$.  
For the three-chain Hubbard model [see Fig.~\ref{fig4}(b)], we again find that the ground 
state in the weak-coupling regime is either a band insulator when $E_g>0$ or an excitonic 
insulator when $E_g < 0$.  The latter includes the Fulde-Ferrell-Larkin-Ovchinnikov 
(FFLO) type of excitonic insulator state when the conduction bands are degenerate, as was 
discussed in Ref.~[\onlinecite{Yamada2016JPSJ}].  In the strong-coupling regime, the 
excitonic insulator state of the BEC type comes out again over a wide region around 
$E_g\simeq 0$.  Thus, it is not surprising that the excitonic insulator state of the BEC type 
appears even when $E_g<0$ if the interaction strength $V$ is sufficiently large.  We have 
argued that Ta$_2$NiSe$_5$ is located in this parameter region.  
Note that the paramagnetic metallic state without excitonic orderings appears for a small 
$V$ region at $E_g<0$ in both the two-chain and three-chain models if the Fermi surface nesting 
is not perfect.

%%%%%%%%%%%%%%%
%\subsection{ARPES double peak and other experiments }
%%%%%%%%%%%%%%%

Finally, let us discuss some other experiments relevant to the present study.  
Most important is the ARPES measurement on Ta$_2$NiSe$_5$ \cite{Seki2014PRB}, which 
indicated that the double-peak structure of the top of the valence band appears over a wide 
temperature range both below and above $T_c$, as shown in Fig.~3  of Ref.~[\onlinecite{Seki2014PRB}].  
This double-peak structure was reproduced by the finite-temperature variational-cluster-approximation 
calculation based on the extended Falicov-Kimball model, as shown in Fig.~4 of 
Ref.~[\onlinecite{Seki2014PRB}], where we note that the noninteracting band structure 
used is semimetallic rather than semiconducting.  This semimetallic band structure, together 
with the strong electron-hole attraction, provided the double-peak structure at the top of the 
valence band.  The band gap remains open at temperatures much higher than $T_c$, so that 
the excitonic condensation is of the BEC type.  The ARPES data on Ta$_2$NiSe$_5$ thus reinforce 
the claim of the present Letter, although such an implication was not emphasized in 
Ref.~[\onlinecite{Seki2014PRB}].  

Another relevant experiment may be a time-resolved ARPES on Ta$_2$NiSe$_5$, where a 
transient metallic phase has been observed by photoexcitation \cite{Ogawa2017}.  
If one supposes that the photoexcited carriers screen the Coulomb interaction $V$ 
in the present excitonic insulator state, without changing other parameters such as the 
band gap, then one may expect that the noninteracting semimetallic band structure should 
be observed (see Fig.~\ref{fig4}), just as the obtained experimental data indicate \cite{Ogawa2017},
although the results depend on the pump fluence \cite{Mor2016}.  Further experimental 
and theoretical studies \cite{Murakami2017} are desirable.  
In this context, the high-pressure experiment \cite{Matsubayashi2017} is also interesting 
if one supposes that the applying pressure enhances the band overlap $E_g$ $(<0)$ 
without changing the Coulomb interaction $V$.  

%%%%%%%%%%%%%%%%%%%
%\subsection{Summary}
%%%%%%%%%%%%%%%%%%%

In summary, we studied the electronic states of an excitonic insulator candidate
Ta$_2$NiSe$_5$ and its related material Ta$_2$NiS$_5$ using the DFT-based electronic
structure calculations and the DMRG and mean-field analyses of the low-energy effective
models, paying particular attention to the observed optical conductivity spectra.
We showed that the spectra in the high-energy regions are described very well by the
DFT-based calculations for both materials, but the description is poor for low energies.
To consider the low-energy spectral features, we assumed an effective model and
calculated the optical conductivity and single-particle spectra using the dynamical 
DMRG method.  We thereby showed that the ground state of Ta$_2$NiSe$_5$ is an excitonic 
insulator of the BEC type, despite the fact that the noninteracting band structure is a 
band-overlap semimetal.  We concomitantly established a novel insulator state caused by 
the strong electron-hole attraction, which is a state of preformed excitons above the transition 
temperature with a finite band gap and may be contrasted with a Mott insulator state caused 
by the strong electron-electron repulsion.  

%\section*{Acknowledgments}

We thank H. Fukuyama, T. Konishi, K. Matsubayashi, T. Mizokawa, C. Monney, H. Okamura, K. Okazaki, H. Sawa, H. Takagi, and T. Toriyama for enlightening discussions.
S. N. acknowledges the technical assistance of U. Nitzsche.
This work was supported by Grants-in-Aid for Scientific Research (Nos.~JP15H06093, JP17K05530, and JP18K13509) from JSPS of Japan and by the SFB 1143 of the Deutsche Forschungsgemeinschaft of Germany.

\end{document}

\title{Supplemental Material for 
\\``Strong Coupling Nature of the Excitonic Insulator State in Ta$_2$NiSe$_5$''}

\author{Koudai Sugimoto$^1$}
\author{Satoshi Nishimoto$^{2,3}$}
\author{Tatsuya Kaneko$^4$}
% \altaffiliation[Also at ]{Physics Department, XYZ University.}%Lines break automatically or can be forced with \\
% \email{sugimoto@chiba-u.jp}
\author{Yukinori Ohta$^5$}%
\affiliation{%
$^1$Center for Frontier Science, Chiba University, Chiba 263-8522, Japan\\
$^2$Institute for Theoretical Solid State Physics, IFW Dresden, 01171 Dresden, Germany\\
$^3$Department of Physics, Technical University Dresden, 01069 Dresden, Germany\\
$^4$Computational Condensed Matter Physics Laboratory, RIKEN, Wako, Saitama 351-0198, Japan\\
$^2$Department of Physics, Chiba University, Chiba 263-8522, Japan
}%

\maketitle

\renewcommand{\theequation}{S.\arabic{equation}}
\renewcommand{\thefigure}{S\arabic{figure}}

In this Supplemental Material, we describe some details of our band structure calculations, 
present the effects of the spin-orbit coupling, and discuss advantages and disadvantages 
of the theoretical models used.  

\section{Details of the band structure calculations}

\begin{figure}[thb]
%\includegraphics[width = 0.9 \linewidth]{bands.eps}
%\begin{overpic}[width = 0.9 \linewidth, grid]{bands.eps}
\begin{overpic}[width = 0.9 \linewidth]{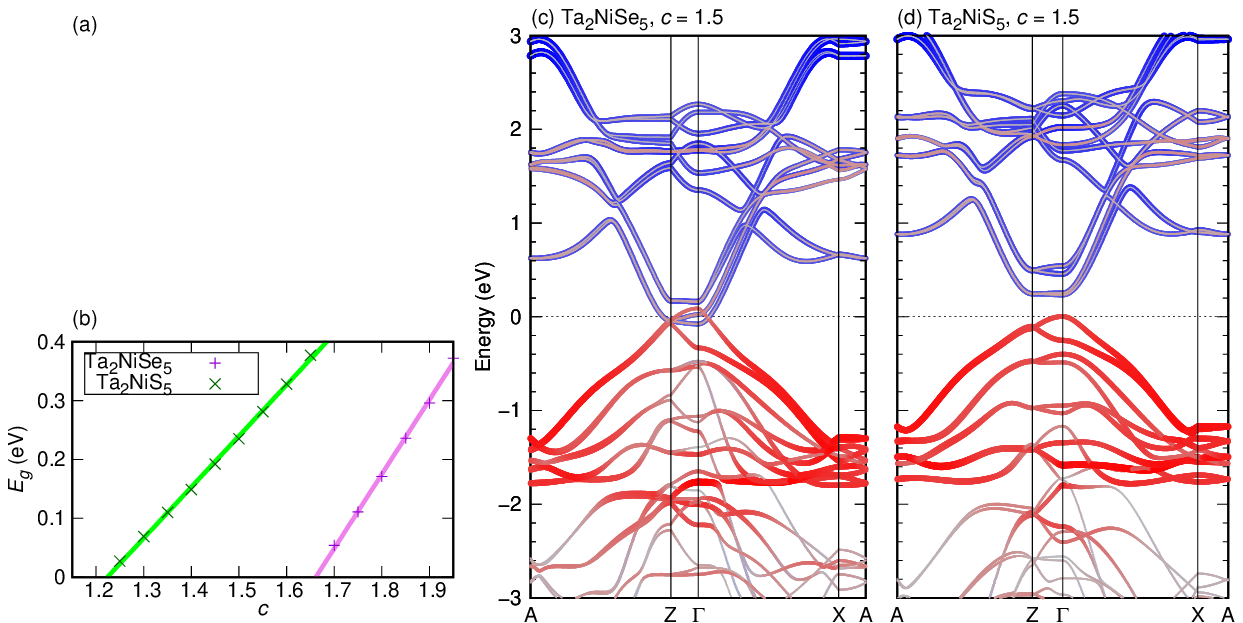}
\put(5,30){\includegraphics[width = 0.25 \linewidth]{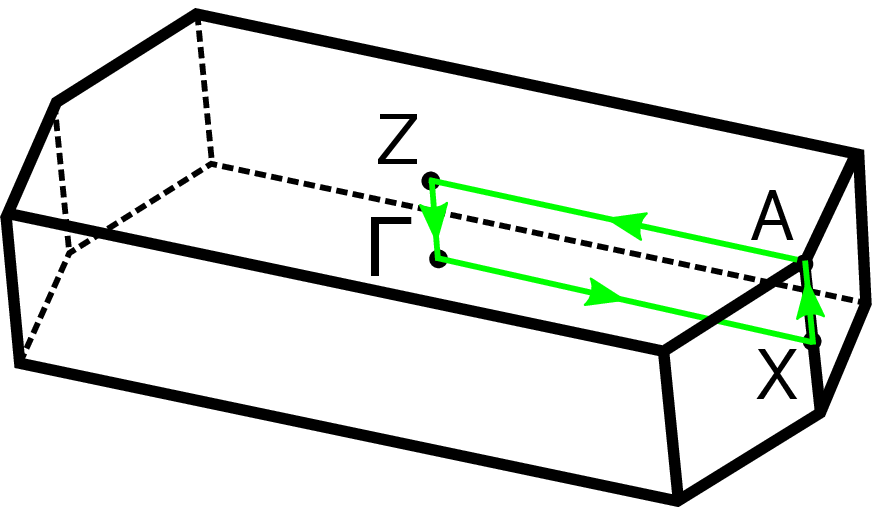}}
\end{overpic}
\caption{
(a) The first Brillouin zone and the path along which the band dispersions are plotted.
(b) Calculated band gaps $E_g$ as a function of the MBJ potential $c$ for 
Ta$_2$NiSe$_5$ (purple) and Ta$_2$NiS$_5$ (green).  The solid lines are the linear least-squares fitting.  
Calculated band dispersions at $c=1.5$ for (c) Ta$_2$NiSe$_5$ and (d) Ta$_2$NiS$_5$, where 
the curve width and color intensity indicate the weight of the Ta $5d$ orbitals (blue) and the Ni $3d$ 
orbitals (red).
}\label{fig:bands}
\end{figure}

We carry out the DFT-based electronic structure calculations using the WIEN2k code [24], 
where we use the generalized gradient approximation for electron correlations with the 
exchange correlation potential given in Ref.~[25].  
The parameters of the crystal structures in the high-temperature orthorhombic phase 
(space group $Cmcm$) with the structure optimization are listed in Table~\ref{tab:crystal} below.  
In the self-consistent calculations, we use 630 $k$-points in the irreducible part of the 
Brillouin zone, assuming the muffin-tin radii $R_{\rm MT}$ of 2.48, 2,27, and 2.16 (1.88) Bohr 
for Ta, Ni, and Se (S), respectively, and the plane-wave cutoff of $K_{\rm max} = 7.0 R_{\rm MT}$.  
The Brillouin zone of the space group $Cmcm$ is shown in Fig.~\ref{fig:bands}(a).  
Because the DFT-based band calculations usually underestimate the band gap $E_g$, we 
introduce a modified Becke-Johnson (MBJ) exchange potential [27] with a parameter $c$ 
defined in Eq.~(3) of Ref.~[23] to improve this underestimation.  
We thus calculate the band gaps of Ta$_2$NiSe$_5$ and Ta$_2$NiS$_5$ as a function of 
$c$ [see Fig.~\ref{fig:bands}(b)], where we find that the band gap opens at $c > 1.2$ for 
Ta$_2$NiS$_5$ but an unusually large value of $c > 1.63$ is required to open the band gap 
for Ta$_2$NiSe$_5$.  We choose a value $c =1.5$ in the main text.  

\begin{table}[h]
\caption{
Left panel: Atomic coordinates of Ta$_2$NiSe$_5$, where the lattice constants are 
$a=3.51232$, $b=13.62805$, and $c=15.78686$ \AA.
Right panel: Atomic coordinates of Ta$_2$NiS$_5$, where the lattice constants are 
$a=3.42997$, $b=13.19963$, and $c=15.21200$ \AA.  
These parameter values are taken from Ref.~[26].}  
\begin{minipage}[t]{0.45\textwidth}
  \begin{center}
    \begin{tabular*}{0.8\textwidth}{@{\extracolsep{\fill}}llll}
      atom & $x$ & $y$ & $z$ \\ \hline
      Ta & 0 & 0.22161 &  0.11071 \\
      Ni & 0 &  0.70234 &  1/4 \\
      Se(1) & 1/2 &  0.08757  &  0.13839 \\
      Se(2) & 0 &  0.15194 &  0.95062 \\
      Se(3) & 0 &  0.32143 & 1/4 
    \end{tabular*}
  \end{center}
\end{minipage}
\begin{minipage}[t]{0.45\textwidth}
  \begin{center}
    \begin{tabular*}{0.8\textwidth}{@{\extracolsep{\fill}}llll}
      atom & $x$ & $y$ & $z$ \\ \hline
      Ta & 0 & 0.27758  & 0.10871  \\
      Ni & 0 & 0.70017 &  1/4 \\
      S(1) & 1/2 &  0.09478  & 0.13549  \\
      S(2) & 0 & 0.15673 & 0.94984 \\
      S(3) & 0 & 0.31188  & 1/4 
    \end{tabular*}
  \end{center}
\end{minipage}
\label{tab:crystal}
\end{table}

The calculated band structures of Ta$_2$NiSe$_5$ and Ta$_2$NiS$_5$ are shown in 
Figs.~\ref{fig:bands}(c) and \ref{fig:bands}(d), respectively.  
In both materials, the conduction bands consist mainly of Ta $5d$ electrons and the 
valence bands consist mainly of Ni $3d$ electrons.  

\begin{figure}[thb]
\includegraphics[width = 0.9\linewidth]{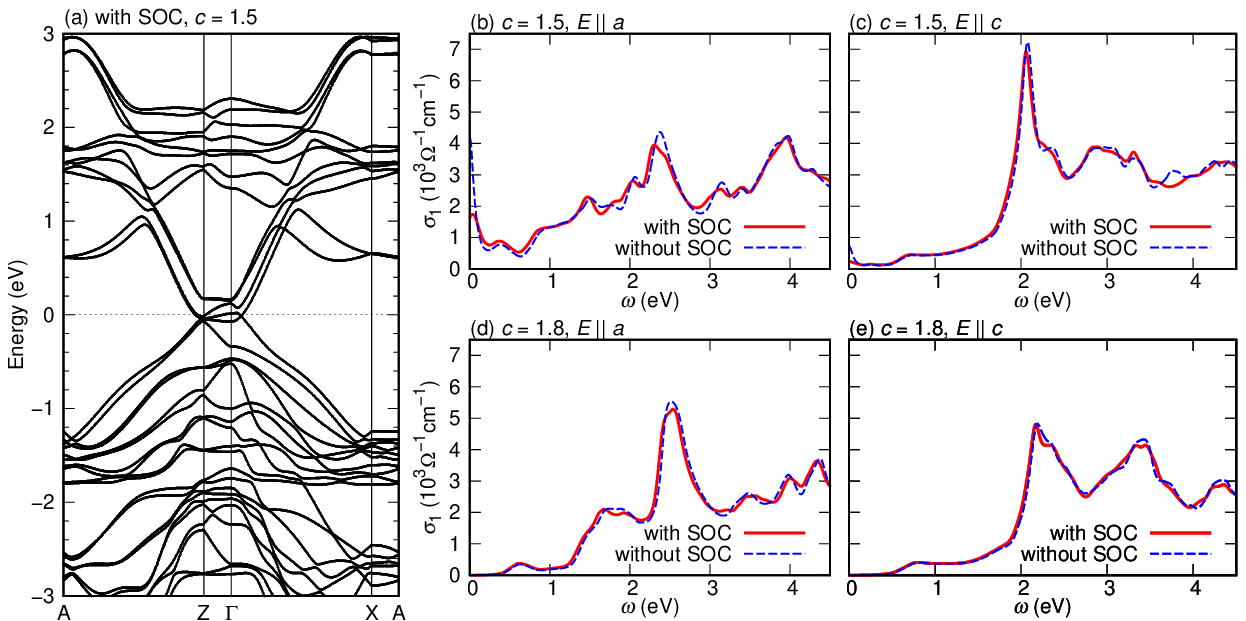}
\caption{
(a) Calculated band dispersions of Ta$_2$NiSe$_5$ at $c = 1.5$, taking into account the SOC.  
Also shown are the calculated optical conductivity spectra of Ta$_2$NiSe$_5$ with and without SOC 
(b) at $c=1.5$ and $E\parallel a$, 
(c) at $c=1.5$ and $E\parallel c$, 
(d) at $c=1.8$ and $E\parallel a$, and 
(e) at $c=1.8$ and $E\parallel c$.  
}\label{fig:SO}
\end{figure}

\section{Effects of spin-orbit coupling}

One might suspect that the band structure is largely affected by the spin-orbit coupling (SOC) 
because the material contains Ta $5d$ electrons.  Here, we investigate the effects of SOC in 
Ta$_2$NiSe$_5$.  The calculations are made by the WIEN2k code and the results for the band 
dispersions are shown in Fig.~\ref{fig:SO}(a).  We find that the degeneracy of the bands is in 
fact lifted along the M-Z line of the Brillouin zone but the effect is not large.  
We also calculate the optical conductivity spectra of Ta$_2$NiSe$_5$.  The results obtained 
with and without SOC are compared in Figs.~\ref{fig:SO}(b)--\ref{fig:SO}(e) for either 
semimetallic ($c=1.5$) or semiconducting situations ($c=1.8$).  We readily find that the effects 
of SOC on the optical conductivity spectra are in general very small.  

\section{Advantages and disadvantages of the theoretical models used}

Besides the DFT-based electronic structure calculations, which are insufficient for describing the 
low-energy electronic states of Ta$_2$NiSe$_5$ as we have shown in the main text, we use three 
simplified theoretical models for discussing the low-energy electronic state of Ta$_2$NiSe$_5$, 
whereby we have provided different levels of description of a complicated many-body problem 
that cannot be solved in general.  

The first one is the 1D extended Falicov-Kimball model (a spinless two-band model), a minimum 
lattice model to discuss the spin-singlet excitonic condensation.  Owing to its simplicity, we can 
apply numerically exact techniques such as the DMRG to the model and discuss essential features 
found experimentally in Ta$_2$NiSe$_5$, such as the band gap opening and large low-energy peak 
in the optical conductivity spectrum [see Fig.~3 in the main text].  

The second one is the two-chain Hubbard model, whereby the spin degrees of freedom of electrons 
and holes are taken into account, so that the spin related physics in the excitonic condensation 
can be discussed.  Also, because the model involves two independent interaction parameters 
(on-site $U$ and inter-chain $V$), we can adjust them to avoid the Hartree shift in the mean-field 
calculation [see Fig.~4(a) in the main text].  

The third one is the three-chain Hubbard model, whereby we can take into account the degeneracy 
of the conduction bands that comes from the three-chain structure of Ta$_2$NiSe$_5$ [10].  
A more realistic calculation of the electronic states of Ta$_2$NiSe$_5$ can in principle be made 
in this model [see Fig.~4(b) in the main text].  

The 2D extended Hubbard model, which we did not refer to in the main text, is also available 
for much more realistic discussions of Ta$_2$NiSe$_5$ [21], for which the theoretical treatment 
becomes more difficult.  One customarily uses the minimum essential model that can reproduce 
the relevant physics of the real material.